\documentclass[showpacs,preprintnumbers,amsmath,amssymb,nofootinbib,floatfix]{revtex4}

\usepackage[dvips]{graphicx}
\usepackage{dcolumn}
\usepackage{bm}

\def\Journal#1#2#3#4{{#1} {\bf #2} (#4) #3}
\def\MPL{Mod. Phys. Lett. A}

\def\NPSUPPL{Nucl. Phys. Proc. Suppl.}
\def\PLB{{Phys. Lett.} B}
\def\PREPO{Phys. Rep.}

\def\PRL{Phys. Rev. Lett.}
\def\RMP{Rev. Mod. Phys.}
\def\PRD{Phys. Rev. D}

\def\PTP{Prog. Theor. Phys.}
\def\JHEP{JHEP}

\def\NPBSUPPL{Nucl. Phys. B. Proc. Suppl.}
\def\EPJ{Euro. Phys. J. C}

\def\JETPUSSR{Sov. Phys. JETP}

\def\ZETP{Zh. Eksp. Teor. Fiz.}

\def\JPG{J. Phys. G}

\def\APJ{Astrophys. J.}

\def\NJP{New J. Phys.}

\def\Erratum{Erratum-ibid}
\def\SCIENCE{Science}

\def\PRAMANA{Pramana J. Phys.}

\begin{document}


 \title{Phases of Flavor Neutrino Masses and CP Violation}

\author{Masaki Yasu\`{e}}
\email{yasue@keyaki.cc.u-tokai.ac.jp}
\affiliation{\vspace{3mm}%
\sl Department of Physics, Tokai University,\\
4-1-1 Kitakaname, Hiratsuka, Kanagawa 259-1292, Japan\\
}


\begin{abstract}
For flavor neutrino masses $M^{PDG}_{ij}$ ($i,j$ = $e,\mu,\tau$) compatible with the phase convention defined by Particle Data Group (PDG), if neutrino mixings are controlled by small corrections to those with $\sin\theta_{13}$=$0$ denoted by $\sin\theta_{13}\delta M^{PDG}_{e\tau}$ and $\sin\theta_{13}\delta M^{PDG}_{\tau\tau}$, CP-violating Dirac phase $\delta_{CP}$ is calculated to be ${\delta _{CP}}$$\approx$$\arg \left[ {\left( M_{\mu \tau }^{PDG \ast}/\tan\theta_{23} + M_{\mu \mu }^{PDG \ast} \right)\delta M_{e\tau }^{PDG} + M_{ee}^{PDG}\delta M_{e\tau }^{PDG \ast} - {\tan\theta_{23}}M_{e\mu }^{PDG}\delta M_{\tau \tau }^{PDG \ast}} \right]$ (mod $\pi$), where $\theta_{ij}$ ($i,j$=1,2,3) denotes an $i$-$j$ neutrino mixing angle. If possible neutrino mass hierarchies are taken into account, the main source of $\delta_{CP}$ turns out to be $\delta M_{e\tau }^{PDG}$ except for the inverted mass hierarchy with ${\tilde m}_1\approx -{\tilde m}_2$, where ${\tilde m}_i=m_ie^{-i\varphi_i}$ ($i$=1,2) stands for a neutrino mass $m_i$ accompanied by a Majorana phase $\varphi_i$ for $\varphi_{1,2,3}$ giving two CP-violating Majorana phases.  We can further derive that ${\delta _{CP}} \approx \arg \left( {M_{e\mu}^{PDG}} \right) - \arg \left( {M_{\mu \mu }^{PDG}} \right)$ with $\arg \left( {M_{e\mu}^{PDG}} \right)\approx \arg \left( {M_{e\tau}^{PDG}} \right)$ for the normal mass hierarchy and ${\delta _{CP}} \approx \arg \left( {M_{ee}^{PDG}} \right) - \arg \left( {M_{e\tau }^{PDG}} \right)+\pi$ for the inverted mass hierarchy with ${\tilde m}_1\approx {\tilde m}_2$.  For specific flavor neutrino masses $M_{ij}$ whose phases arise from $M_{e\mu,e\tau,\tau\tau}$, these phases can be connected with $\arg (M_{ij}^{PDG})$ ($i,j$=$e,\mu,\tau$).  As a result, numerical analysis suggests that Dirac CP-violation becomes maximal as $\left|\arg(M_{e\mu})\right|$ approaches to $\pi/2$ for the inverted mass hierarchy with ${\tilde m}_1\approx {\tilde m}_2$ and for the degenerate mass pattern satisfying the inverted mass ordering and that Majorana CP-violation becomes maximal as $\left|\arg\left( M_{\tau\tau}\right)\right|$ approaches to its maximal value around $0.5$ for the normal mass hierarchy.  Alternative CP-violation induced by three CP-violating Dirac phases is compared with the conventional one induced by $\delta_{CP}$ and two CP-violating Majorana phases.
\end{abstract}

\maketitle

Various experimental evidences of neutrino oscillations provided by the atmospheric \cite{atmospheric}, solar \cite{oldsolar,solar}, reactor \cite{reactor, theta13} and accelerator \cite{accelerator} neutrino oscillation experiments have indicated that neutrinos have tiny masses and their flavor states are mixed with each other. Nowadays, to study CP violation in neutrinos is one of the important issues to be addressed in order to further understand neutrino physics. The recent observation on the nonvanishing reactor neutrino mixing \cite{theta13} has opened the possibility that details of Dirac CP-violation can be experimentally clarified in near future.  Theoretically, effects of CP-violation are described in terms of three phases, one CP-violating Dirac phase $\delta_{CP}$ and two CP-violating Majorana phases $\phi_{2,3}$ \cite{CPViolation}.  Neutrino mixings are parameterized by the Pontecorvo-Maki-Nakagawa-Sakata (PMNS) unitary matrix $U_{PMNS}$ \cite{PMNS}, which converts the massive neutrinos $\nu_i$ ($i=1,2,3$) into the flavor neutrinos $\nu_f$ ($f=e,\mu,\tau$).  The standard description of $U_{PMNS}$ adopted by the Particle Data Group (PDG) \cite{PDG} is given by $U^{PDG}_{PMNS}=U^0_\nu K^0$ with
\begin{eqnarray}
U^0_\nu&=&\left( \begin{array}{ccc}
  c_{12}c_{13} &  s_{12}c_{13}&  s_{13}e^{-i\delta_{CP}}\\
  -c_{23}s_{12}-s_{23}c_{12}s_{13}e^{i\delta_{CP}}
                                 &  c_{23}c_{12}-s_{23}s_{12}s_{13}e^{i\delta_{CP}}
                                 &  s_{23}c_{13}\\
  s_{23}s_{12}-c_{23}c_{12}s_{13}e^{i\delta_{CP}}
                                 &  -s_{23}c_{12}-c_{23}s_{12}s_{13}e^{i\delta_{CP}}
                                 & c_{23}c_{13}\\
\end{array} \right),
\nonumber \\
K^0 &=& {\rm diag}(1, e^{i\phi_2/2}, e^{i\phi_3/2}),
\label{Eq:UuPDG}
\end{eqnarray}
where $c_{ij}=\cos\theta_{ij}$ and $s_{ij}=\sin\theta_{ij}$ with $\theta_{ij}$ representing a $\nu_i$-$\nu_j$ mixing angle ($i,j$=1,2,3). The best fit values of the observed results in the case of the normal mass ordering are summarized as \cite{NuData}:  
\begin{eqnarray}
\Delta m^2_{21} ~[10^{-5}~{\rm eV}^2] & = &7.62\pm 0.19,
\quad
\Delta m^2_{31} ~[10^{-3}~{\rm eV}^2] = 2.55
{\footnotesize
{\begin{array}{*{20}c}
   { + 0.06}  \\
   { - 0.09}  \\
\end{array}}
},
\label{Eq:NuDataMass}\\
\sin ^2 \theta _{12} & = &0.320
{\footnotesize
{\begin{array}{*{20}c}
   { + 0.016}  \\
   { - 0.017}  \\
\end{array}}
},
\quad
\sin ^2 \theta _{23} =0.427
{\footnotesize
{\begin{array}{*{20}c}
   { + 0.034}  \\
   { - 0.027}  \\
\end{array}}
}
~\left(
0.613
{\footnotesize
{\begin{array}{*{20}c}
   { + 0.022}  \\
   { - 0.040}  \\
\end{array}}
}
\right),
\quad
\sin ^2 \theta _{13} =0.0246
{\footnotesize
 {\begin{array}{*{20}c}
   { + 0.0029}  \\
   { - 0.0028}  \\
\end{array}}
},
\label{Eq:NuDataAngle}\\
\frac{\delta _{PC}}{\pi} &=& 0.80
{\footnotesize
 {\begin{array}{*{20}c}
   { + 1.20}  \\
   { - 0.80}  \\
\end{array}}
},
\label{Eq:NuDataPhase}
\end{eqnarray}
where $\Delta m^2_{ij}=m^2_i-m^2_j$ with $m_i$ representing a mass of $\nu_i$ ($i=1,2,3$).  The quoted values in the case of the inverted mass ordering ($\Delta m^2_{31}<0$) are not so different from Eqs.(\ref{Eq:NuDataMass})-(\ref{Eq:NuDataPhase}).  There is another similar analysis with $\Delta m^2_{23}$ defined as $\Delta m^2_{23}=m^2_3-(m^2_1+m^2_2)/2$ that has reported the slightly smaller values of $\sin^2\theta_{23} = 0.365-0.410$ \cite{NuData2}.

In this note, we would like to address the issue of leptonic CP-violation with the emphasis laid on the role of phases of flavor neutrino masses and to find possible correlations between phases of flavor neutrino masses and $\delta_{CP}$ and $\phi_{2,3}$ of CP-violation.  CP-violating phases arise from complex flavor neutrino masses.  However, because of the freedom of choosing charged-lepton phases, phases of neutrino masses are not uniquely defined.  Namely, different phase structure gives the same effects of CP-violation.  We first discuss how to relate phases of flavor neutrino masses to observed quantities.  To do so, we use a neutrino mass matrix $M^{PDG}$, whose phases are so chosen that the corresponding eigenvectors giving $U_{PMNS}$ show the phase convention defined by PDG, which is nothing but Eq.(\ref{Eq:UuPDG}). Next, we give theoretical and numerical estimation of phases of flavor neutrino masses and present possible correlations with CP-violating phases. Also discussed is alternative CP-violation characterized by three CP-violating Dirac phases \cite{ThreeDiracPhases}, which has an advantage to discuss property of neutrinoless double beta decay \cite{DoubleBeta}. 

We start with discussions based on $M^{PDG}$ defined to be:
\begin{eqnarray}
&&
M^{PDG} = \left( \begin{array}{*{20}{c}}
M_{ee}^{PDG}&M_{e\mu }^{PDG}&M_{e\tau }^{PDG}\\
M_{e\mu }^{PDG}&M_{\mu \mu }^{PDG}&M_{\mu \tau }^{PDG}\\
M_{e\tau }^{PDG}&M_{\mu \tau }^{PDG}&M_{\tau \tau }^{PDG}
\end{array} \right).
\label{Eq:M_PDG}
\end{eqnarray}
Since $\delta_{CP}$ is associated with $\sin\theta_{13}$, it is useful to divide $M^{PDG}$ into two pieces consisting of $M^{PDG}_{\theta_{13}=0}$ giving $\sin\theta_{13}=0$ and $\Delta M^{PDG}$ inducing $\sin\theta_{13}\neq 0$ \cite{theta_13=0}:
\begin{equation}
M^{PDG} = M^{PDG}_{\theta_{13}=0}+\Delta {M^{PDG}},
\label{Eq:Msin13Delta}
\end{equation}
with
\begin{eqnarray}
&&
M_{{\theta _{13}} = 0}^{PDG} = \left( {\begin{array}{*{20}{c}}
{{M_{ee}^{PDG}}}&{{M_{e\mu }^{PDG}}}&{ - {t_{23}}{M_{e\mu }^{PDG}}}\\
{{M_{e\mu }^{PDG}}}&{{M_{\mu \mu }^{PDG}}}&{{M_{\mu \tau }^{PDG}}}\\
{ - {t_{23}}{M_{e\mu }^{PDG}}}&{{M_{\mu \tau }^{PDG}}}&{{M_{\mu \mu }^{PDG}} + \frac{{1 - t_{23}^2}}{{{t_{23}}}}{M_{\mu \tau }^{PDG}}}
\end{array}} \right),
\nonumber\\
&&
\Delta {M^{PDG}} = \left( {\begin{array}{*{20}{c}}
0&0&{{M_{e\tau }^{PDG}} + {t_{23}}{M_{e\mu }^{PDG}}}\\
0&0&0\\
{{M_{e\tau }^{PDG}} + {t_{23}}{M_{e\mu }^{PDG}}}&0&{{M_{\tau \tau }^{PDG}} - \left( {{M_{\mu \mu }^{PDG}} + \frac{{1 - t_{23}^2}}{{{t_{23}}}}{M_{\mu \tau }^{PDG}}} \right)}
\end{array}} \right).
\label{Eq:Msin13zero}
\end{eqnarray}
It should be noted that Eq.(\ref{Eq:Msin13Delta}) is just an identity. There are specific models giving $M_{{\theta _{13}} = 0}^{PDG}$ \cite{Tribimaximal, Scaling, Bipair, Others}, whose predictions on CP-violation can be covered by our discussions.  

Noticing that $M^{PDG} = U_{PMNS}^\ast M_{mass}U_{PMNS}^\dagger$, where $M_{mass}$ = diag.$(m_1, m_2, m_3)$, we can express $M^{PDG}_{ij}$ in terms of masses, mixing angles and phases including three Majorana phases $\varphi_{1,2,3}$ that gives $\phi_i=\varphi_i-\varphi_1$.  Since $\sin\theta_{13}$ acts as a correction parameter, $\Delta {M^{PDG}}$ is redefined to be $\sin\theta_{13}\delta {M^{PDG}}$:
\begin{eqnarray}
\sin\theta_{13}\delta M^{PDG}_{e\tau} &=& {M_{e\tau }^{PDG}} + {t_{23}}{M_{e\mu }^{PDG}},
\nonumber\\
\sin\theta_{13}\delta M^{PDG}_{\tau\tau} &=& {M_{\tau \tau }^{PDG}} - \left( {{M_{\mu \mu }^{PDG}} + \frac{{1 - t_{23}^2}}{{{t_{23}}}}{M_{\mu \tau }^{PDG}}} \right),
\label{Eq:deltaM_formula}
\end{eqnarray}
from which $\delta M^{PDG}_{e\tau}$ and $\delta M^{PDG}_{\tau\tau}$ are calculated to be:
\begin{eqnarray}
\delta M^{PDG}_{e\tau} &=& \frac{{c_{13}}}{{c_{23}}}\left[ {{e^{ i{\delta _{CP}}}}{{\tilde m}_3} - {e^{-i{\delta _{CP}}}}\left( {c_{12}^2{{\tilde m}_1} + s_{12}^2{{\tilde m}_2}} \right)} \right],
\nonumber\\
\delta M^{PDG}_{\tau\tau} &=& \frac{{c_{12}}{s_{12}}}{{s_{23}}{c_{23}}}{e^{-i{\delta _{CP}}}}\left( {{{\tilde m}_2} - {{\tilde m}_1}} \right),
\label{Eq:results}
\end{eqnarray}
where ${\tilde m}_i= m_i e^{-i\varphi_i}$ ($i=1,2,3$).
To estimate CP-violating Dirac phase, let us consider ${\bf M} = M^{PDG\dagger}M^{PDG}$.  The quantity of $s_{23}{\bf M}_{e\mu} + c_{23}{\bf M}_{e\tau}$ corresponding to $\Delta M^{PDG}_{e\tau}(={M_{e\tau }^{PDG}} + {t_{23}}{M_{e\mu }^{PDG}})$ is also known to vanish at $\theta_{13}=0$ \cite{DiracVsMsss}.  In fact, it is expressed in terms of observed masses and mixing angles to be:
\begin{eqnarray}
{s_{23}}{{\bf M}_{e\mu }} + {c_{23}}{{\bf M}_{e\tau }} &=& {c_{13}}{s_{13}}{e^{ - i{\delta _{CP}}}}\left[ {m_3^2 - \left( {c_{12}^2m_1^2 + s_{12}^2m_2^2} \right)} \right].
\label{Eq:masssquared}
\end{eqnarray}
On the other hand, Eq.(\ref{Eq:Msin13Delta}) yields
\begin{eqnarray}
{s_{23}}{{\bf M}_{e\mu }} + {c_{23}}{{\bf M}_{e\tau }} &=& 
s_{13}c_{23}^2\left[ {\left( {\frac{1}{{{t_{23}}}}M_{\mu \tau }^{PDG} + M_{\mu \mu }^{PDG} + s_{13}\delta M_{\tau \tau }^{PDG}} \right)\delta M_{e\tau }^{PDG\ast }} \right.
\nonumber\\
&& + \left. {M_{ee}^{PDG\ast }\delta M_{e\tau }^{PDG} - {t_{23}}M_{e\mu }^{PDG\ast }\delta M_{\tau \tau }^{PDG}} \right].
\label{Eq:deltaCP_masss_relation}
\end{eqnarray}
Since $s_{13}\delta M_{\tau \tau }^{PDG}\delta M_{e\tau }^{PDG\ast }$ in Eq.(\ref{Eq:deltaCP_masss_relation}) can be safely neglected, CP-violating Dirac phase $\delta _{CP}$ is approximated to be:
\begin{eqnarray}
{\delta _{CP}} &\approx&  \arg \left[ \left( \frac{1}{{{t_{23}}}}M_{\mu \tau }^{PDG\ast} + M_{\mu \mu }^{PDG \ast} \right)\delta M_{e\tau }^{PDG} + M_{ee}^{PDG}\delta M_{e\tau }^{PDG \ast} - t_{23}M_{e\mu }^{PDG}\delta M_{\tau \tau }^{PDG \ast} \right],
\label{Eq:deltaCP_masssquared}
\end{eqnarray}
where an extra $\pi$ should be added to $\delta _{CP}$ if $m_3^2 - \left(c_{12}^2m_1^2 + s_{12}^2m_2^2\right) < 0$.

To discus more about $\delta_{CP}$, since contributions of flavor neutrino masses to $\delta_{CP}$ depend on their magnitudes, we may include various constraints on $M^{PDG}_{ij}$ supplied by mass hierarchies: $m^2_{1,2,3}$: $m^2_1 < m^2_2 \ll m^2_3$ as normal mass hierarchy, $m^2_3 \ll m^2_1 < m^2_2$ as inverted mass hierarchy and $m^2_1 < m^2_2 \sim m^2_3$ as degenerate mass pattern with $m^2_1 < m^2_2 \approx m^2_3$ (or $m^2_3\approx m^2_1 < m^2_2$).  The magnitudes of masses are controlled by the ideal case of $\theta_{13}=0$ since corrections to the ideal case are $\mathcal{O}(\sin^2\theta_{13})$ \cite{FukiYasue}.  For $\theta_{13}=0$, we have the following estimates of three masses and two mixing angles:
\begin{eqnarray}
{\tilde m}_1 &=& \frac{{c_{12}^2{M^{PDG}_{ee}} - s_{12}^2\left( {{M^{PDG}_{\mu \mu }} - {t_{23}}{M^{PDG}_{\mu \tau }}} \right)}}{{c_{12}^2 - s_{12}^2}}
=
\frac{{{M^{PDG}_{ee}} + {M^{PDG}_{\mu \mu }} - {t_{23}}{M^{PDG}_{\mu \tau }}}}{2} - \frac{{{M^{PDG}_{e\mu }}}}{{{c_{23}}\sin 2{\theta _{12}}}},
\nonumber\\
{\tilde m}_2 &=& \frac{{c_{12}^2\left( {{M^{PDG}_{\mu \mu }} - {t_{23}}{M^{PDG}_{\mu \tau }}} \right) - s_{12}^2{M^{PDG}_{ee}}}}{{c_{12}^2 - s_{12}^2}}
=
\frac{{{M^{PDG}_{ee}} + {M^{PDG}_{\mu \mu }} - {t_{23}}{M^{PDG}_{\mu \tau }}}}{2} + \frac{{{M^{PDG}_{e\mu }}}}{{{c_{23}}\sin 2{\theta _{12}}}},
\label{mass_13=0}\\
{\tilde m}_3 &=& {M^{PDG}_{\mu \mu }} + \frac{1}{{{t_{23}}}}{M^{PDG}_{\mu \tau }},
\nonumber
\end{eqnarray}
and
\begin{equation}
\tan {\theta _{23}} =  - \frac{{{M^{PDG}_{e\tau }}}}{{{M^{PDG}_{e\mu }}}},
\quad
\tan 2{\theta _{12}} = \frac{2}{{{c_{23}}}}\frac{{{M^{PDG}_{e\mu }}}}{{{M^{PDG}_{\mu \mu }} - {t_{23}}{M^{PDG}_{\mu \tau }} - {M^{PDG}_{ee}}}}.
\label{angle_13=0}
\end{equation}
We are, then, allowed to use the following gross structure of $M_{{\theta _{13}} = 0}^{PDG}$ \cite{IdealMassMatrix}:
\begin{equation}
M_{{\theta _{13}} = 0}^{PDG} \approx 
\left( {\begin{array}{*{20}{c}}
0&0&0\\
0&1&1/t_{23}\\
0&1/t_{23}&1/t_{23}^2
\end{array}} \right){M^{PDG}_{\mu \mu }},
\label{M_angle_13=0_normal}
\end{equation}
for the normal mass hierarchy (NMH) \cite{theta_13=0}, and
\begin{equation}
M_{{\theta _{13}} = 0}^{PDG} \approx 
\left( {\begin{array}{*{20}{c}}
2&0&0\\
0&1&{ - {t_{23}}}\\
0&{ - {t_{23}}}&{t_{23}^2}
\end{array}} \right){M^{PDG}_{\mu \mu }},
\label{M_angle_13=0_inverted1}
\end{equation}
for the inverted mass hierarchy with ${\tilde m}_1 \approx {\tilde m}_2$ (IMH-1) \cite{Scaling}, and
\begin{equation}
M_{{\theta _{13}} = 0}^{PDG} \approx 
\left( {\begin{array}{*{20}{c}}
{ - 2}&{ - 2{c_{23}}\tan 2{\theta _{12}}}&{2{s_{23}}\tan2{\theta _{12}}}\\
{ - 2{c_{23}}\tan 2{\theta _{12}}}&1&{ - {t_{23}}}\\
{2{c_{23}}\tan 2{\theta _{12}}}&{ - {t_{23}}}&{t_{23}^2}
\end{array}} \right){M^{PDG}_{\mu \mu }},
\label{M_angle_13=0_inverted2}
\end{equation}
for the inverted mass hierarchy with ${\tilde m}_1\approx -{\tilde m}_2$ (IMH-2) \cite{m1=-m2}, and
\begin{equation}
M_{{\theta _{13}} = 0}^{PDG} \approx 
\left( {\begin{array}{*{20}{c}}
1&0&0\\
0&{\cos 2{\theta _{23}}}&{ - \sin 2{\theta _{23}}}\\
0&{ - \sin 2{\theta _{23}}}&{ - \cos 2{\theta _{23}}}
\end{array}} \right)M_{ee}^{PDG},
\label{M_angle_13=0_degenerate}
\end{equation}
for the degenerate mass pattern with ${\tilde m}_1\approx {\tilde m}_2 \approx -{\tilde m}_3$ (DMP) \cite{DegenerateNeutrino}.\footnote{Since $M^{PDG}_{\mu\tau}$ does not vanish in the limit of ${\tilde m}_1 = {\tilde m}_2 = {\tilde m}_3$ because of the presence of $s_{13}e^{i\delta_{CP}}$, the simplest case of ${\tilde m}_1 \approx {\tilde m}_2 \approx {\tilde m}_3$ requiring fairly suppressed magnitude of  $M^{PDG}_{\mu\tau}$ is not relevant. In other cases with ${\tilde m}_1 \approx -{\tilde m}_2$, relations among masses are complicated and seem to give no positive feedback to our discussions.}

\begin{figure}[t]
\begin{center}
\includegraphics*[10mm,226mm][300mm,290mm]{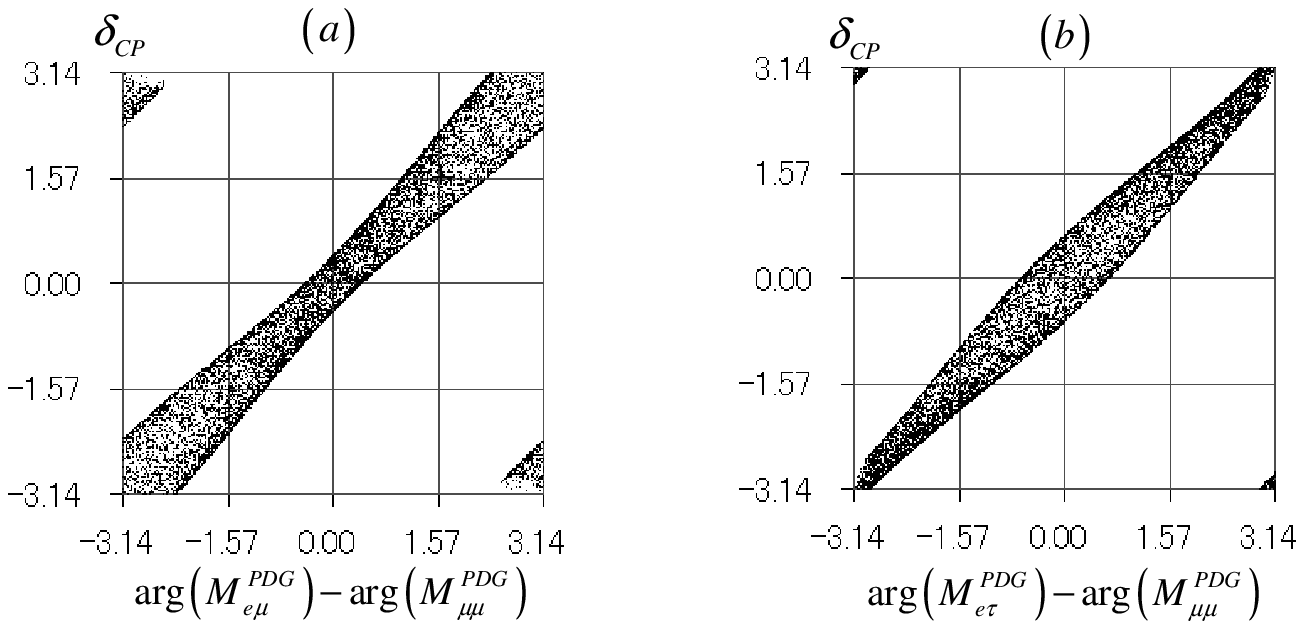}
\caption{The predictions of $\delta_{CP}$ for the normal mass hierarchy (NMH): (a) $\delta_{CP}\approx\arg \left( {M_{e\mu }^{PDG}} \right) - \arg \left( {M_{\mu \mu }^{PDG}} \right)$ or (b) $\delta_{CP}\approx\arg \left( {M_{e\tau }^{PDG}} \right) - \arg \left( {M_{\mu \mu }^{PDG}} \right)$.}
\label{Fig:Simplified_normal}
\includegraphics*[10mm,226mm][300mm,290mm]{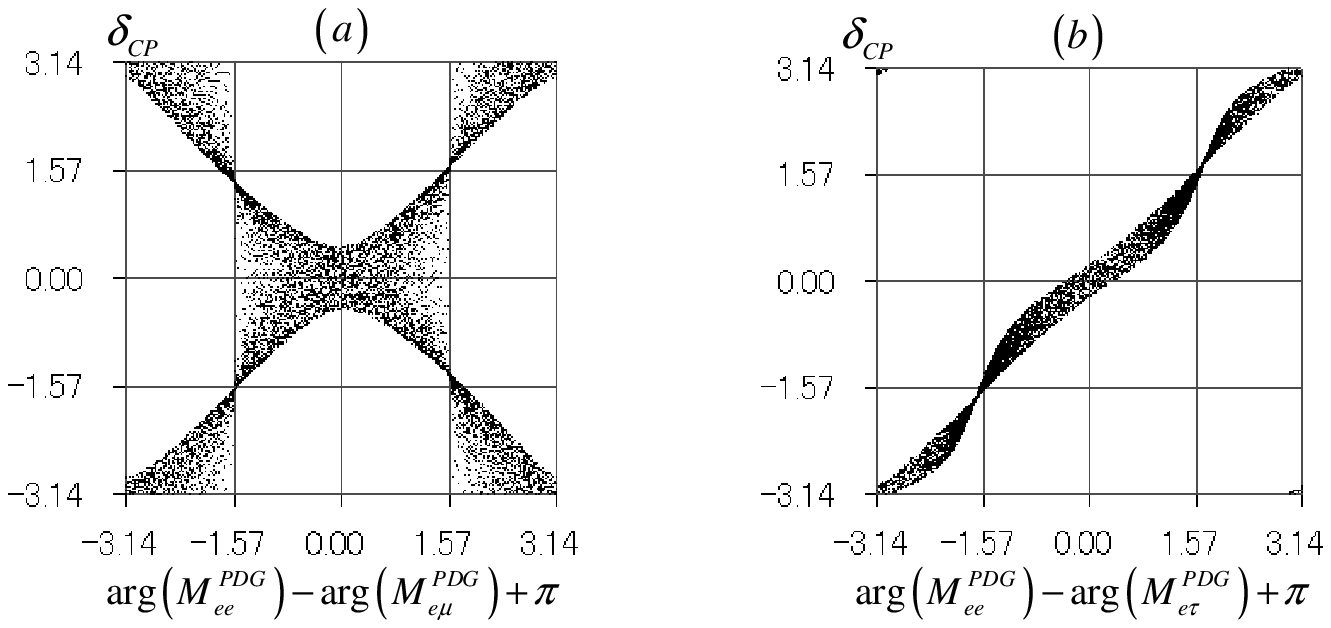}
\caption{The predictions of $\delta_{CP}$ for the inverted mass hierarchy with ${\tilde m}_1\approx {\tilde m}_2$ (IMH-1): (a) $\delta_{CP}\approx\arg \left( {M_{ee }^{PDG}} \right) - \arg \left( {M_{e\mu }^{PDG}} \right)+\pi$ or (b) $\delta_{CP}\approx\arg \left( {M_{ee }^{PDG}} \right) - \arg \left( {M_{e\tau }^{PDG}} \right)+\pi$.}
\label{Fig:Simplified_inverted}
\end{center}
\end{figure}

Applying these estimates to Eq.(\ref{Eq:deltaCP_masssquared}), we reach 
\begin{enumerate}
\item for NMH, ignoring $M^{PDG}_{ee,e\mu,e\tau}$,
\begin{equation}
{\delta _{CP}} \approx \arg \left( {M_{\mu \mu }^{PDG\ast }\delta M_{e\tau }^{PDG}} \right),
\label{Eq:deltaCP_normal}
\end{equation}
\item for IMH-1, ignoring $M^{PDG}_{e\mu,e\tau}$,
\begin{equation}
{\delta _{CP}} \approx \arg \left( {M_{ee}^{PDG}\delta M_{e\tau }^{PDG\ast }} \right)+\pi,
\label{Eq:deltaCP_inverted1}
\end{equation}
\item for IMH-2, ignoring $M^{PDG}_{\mu\mu,\mu\tau,\tau\tau}$,
\begin{equation}
{\delta _{CP}} \approx \arg \left( {M_{ee}^{PDG}\delta M_{e\tau }^{PDG\ast } - {t_{23}}M_{e\mu }^{PDG}\delta M_{\tau \tau }^{PDG\ast }} \right)+\pi,
\label{Eq:deltaCP_inveted2}
\end{equation}
\item for DMP, ignoring $M^{PDG}_{e\mu,e\tau}$,
\begin{equation}
{\delta _{CP}} \approx \arg \left[ {\left( {\frac{1}{{{t_{23}}}}M_{\mu \tau }^{PDG\ast } + M_{\mu \mu }^{PDG\ast }} \right)\delta M_{e\tau }^{PDG} + M_{ee}^{PDG}\delta M_{e\tau }^{PDG\ast }} \right]\left( +\pi\right),
\label{Eq:deltaCP_degenerate}
\end{equation}
with an extra $\pi$ for the inverted mass ordering.
\end{enumerate}
It is thus concluded that the main source of $\delta_{CP}$ is $\delta M_{e\tau }^{PDG}$ except for IMH-2.  This conclusion is in accord with the expectation from Eq.(\ref{Eq:results}) that $\delta M_{\tau\tau}^{PDG}$ is suppressed unless ${\tilde m}_1 \approx -{\tilde m}_2$ as in IMH-2.
Since $\arg\left( M_{e\mu}^{PDG}\right)= \arg\left( M_{e\tau}^{PDG}\right)$ is valid for $\sin\theta_{13}=0$, we expect that $\arg\left( M_{e\mu}^{PDG}\right)\approx \arg\left( M_{e\tau}^{PDG}\right)$ is preserved for $\sin\theta_{13}\neq 0$ especially in NMH because the single term proportional to ${\tilde m}_3$ will dominate in $M_{e\mu, e\tau}^{PDG}$.  Using the approximation of $\arg \left( {\delta M_{e\tau }^{PDG}} \right) = \arg \left( {M_{e\tau }^{PDG} + {t_{23}}M_{e\mu }^{PDG}} \right)\approx\arg \left( M_{e\mu, e\tau }^{PDG}\right)$, we can find more simplified relation from Eq.(\ref{Eq:deltaCP_normal}) in NMH:
\begin{equation}
{\delta _{CP}} \approx \arg \left( {M_{e\mu}^{PDG}} \right) - \arg \left( {M_{\mu \mu }^{PDG}} \right),
\label{Eq:deltaCP_PDG_normal}
\end{equation}
with $\arg \left( {M_{e\mu}^{PDG}} \right)\approx \arg \left( {M_{e\tau}^{PDG}} \right)$.  The similar relation is also found for IMH-1 and dictates from Eq.(\ref{Eq:deltaCP_inverted1}) that
\begin{equation}
{\delta _{CP}} \approx \arg \left( {M_{ee}^{PDG}} \right) - \arg \left( {M_{e\tau }^{PDG}} \right) + \pi,
\label{Eq:deltaCP_PDG_inverted}
\end{equation}
where $\arg \left( {M_{e\mu}^{PDG}} \right)\approx \arg \left( {M_{e\tau}^{PDG}} \right)$ does not serve as a good approximation. In fact, Eq.(\ref{Eq:deltaCP_PDG_inverted}) using another choice of $\arg \left( {M_{e\mu }^{PDG}} \right)$ instead of $\arg \left( {M_{e\tau }^{PDG}} \right)$ is not numerically supported (See FIG.\ref{Fig:Simplified_inverted}-(a)).

To further enhance predictability based on our approach to CP-violations, we have to minimize the number of phases present in flavor neutrino masses, which can be as small as three.  Therefore, a plausible program to discuss linkage between CP-violating phases and flavor neutrino masses is 
\begin{enumerate}
\item to construct a reference mass matrix to be denoted by $M_\nu$ with unique choice of phases of neutrino masses, 
\item to construct a general mass matrix to be denoted by $M$ that includes the ambiguity of charged-lepton phases to cover all phase structure, which is linked to $M_\nu$,
\item to construct $M^{PDG}$ converted from $M$, whose eigenvectors yield $U^{PDG}_{PMNS}$.
\end{enumerate}
Since flavor neutrino masses in $M^{PDG}$ are expressed by measured quantities, useful information on phases of $M_\nu$ can be extracted from $M^{PDG}$.
\begin{figure}[t]
\begin{center}
\includegraphics*[10mm,226mm][300mm,290mm]{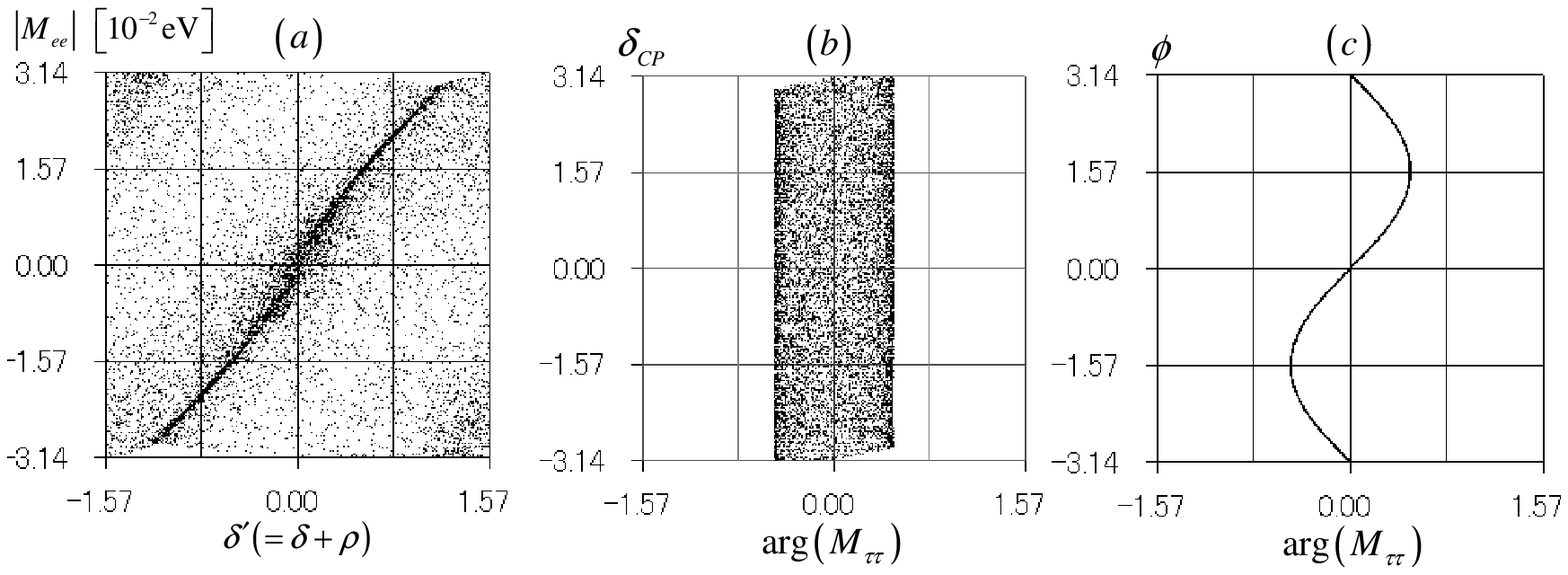}
\caption{The predictions of (a) $\delta_{CP}$ as a function of $\arg\left( M_{e\mu}\right)$, (b) $\delta_{CP}$ as a function of $\arg\left( M_{\tau\tau}\right)$ and (c) $\phi$ (=$\varphi_3-\varphi_2$) as a function of $\arg\left( M_{\tau\tau}\right)$ for the normal mass hierarchy (NMH).}
\label{Fig:Normal}
\includegraphics*[10mm,226mm][300mm,290mm]{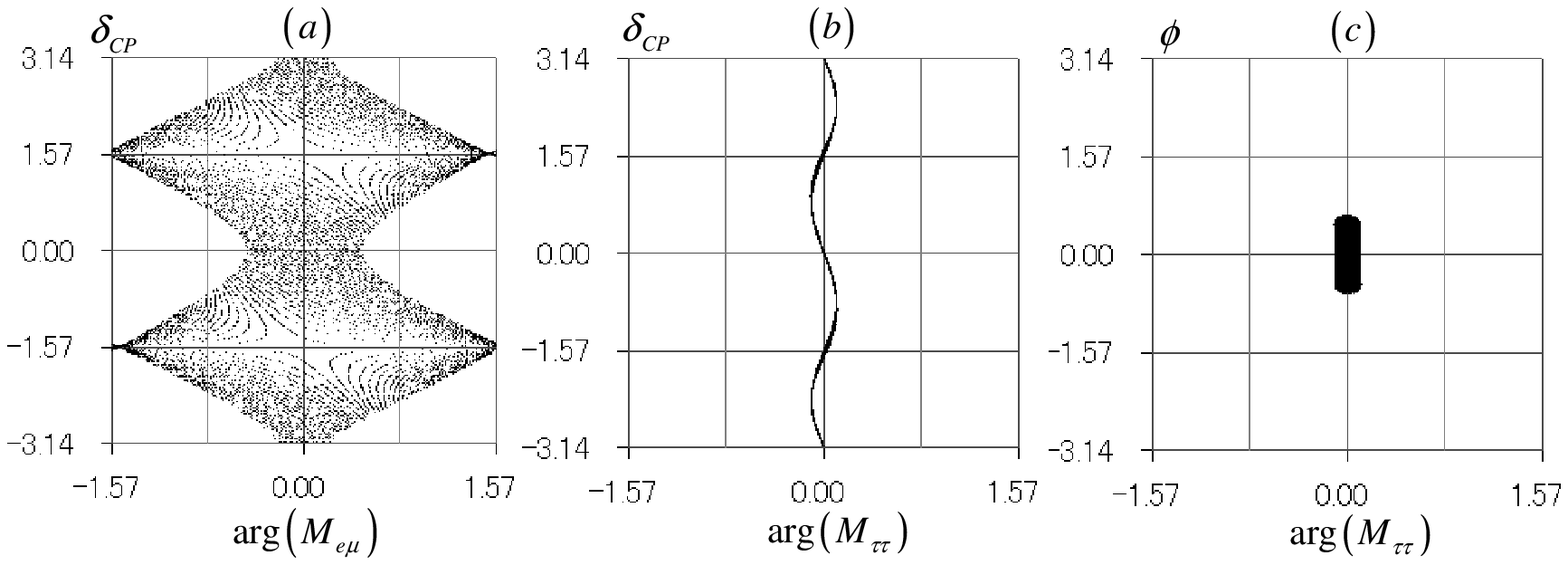}
\caption{The same as in FIG.\ref{Fig:Normal} but for the inverted mass hierarchy with ${\tilde m}_1\approx {\tilde m}_2$ (IMH-1) and $\phi$ = $\varphi_2-\varphi_1$.}
\label{Fig:Inverted1}
\end{center}
\end{figure}
\begin{figure}[t]
\begin{center}
\includegraphics*[10mm,226mm][300mm,290mm]{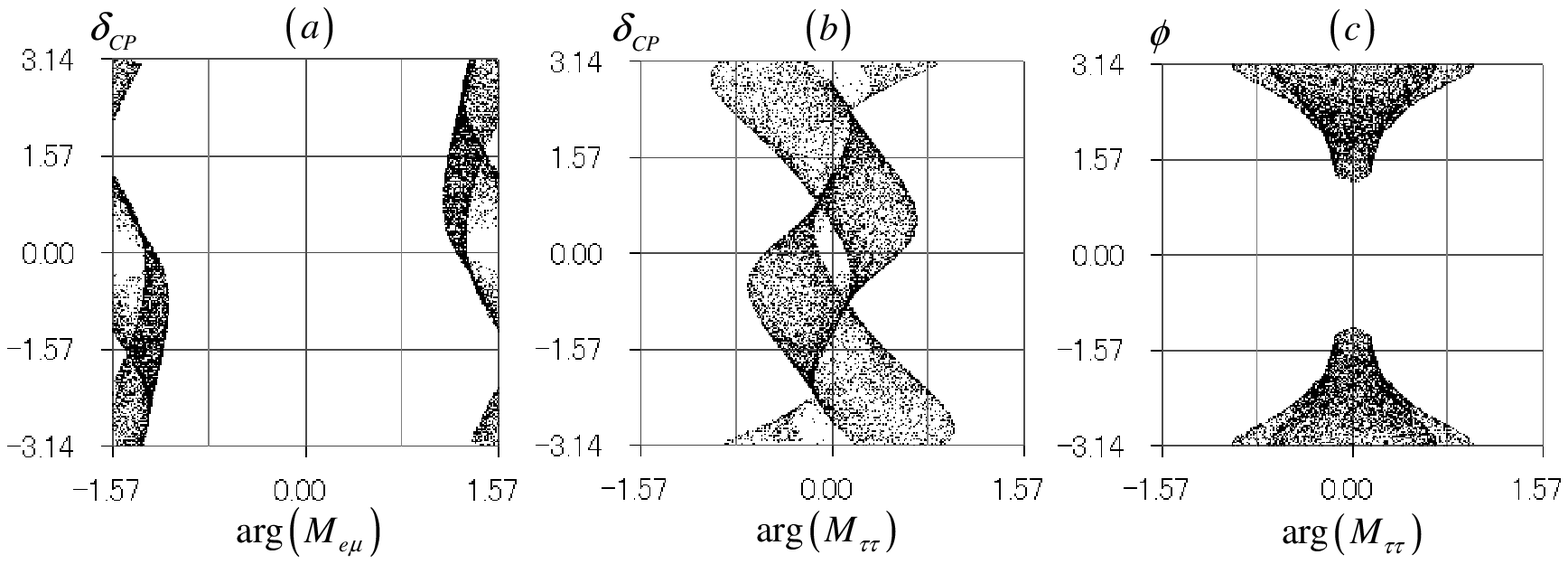}
\caption{The same as in FIG.\ref{Fig:Normal} but for the inverted mass hierarchy with ${\tilde m}_1\approx -{\tilde m}_2$ (IMH-2).}
\label{Fig:Inverted2}
\end{center}
\end{figure}
\begin{figure}[t]
\begin{center}
\includegraphics*[10mm,226mm][300mm,290mm]{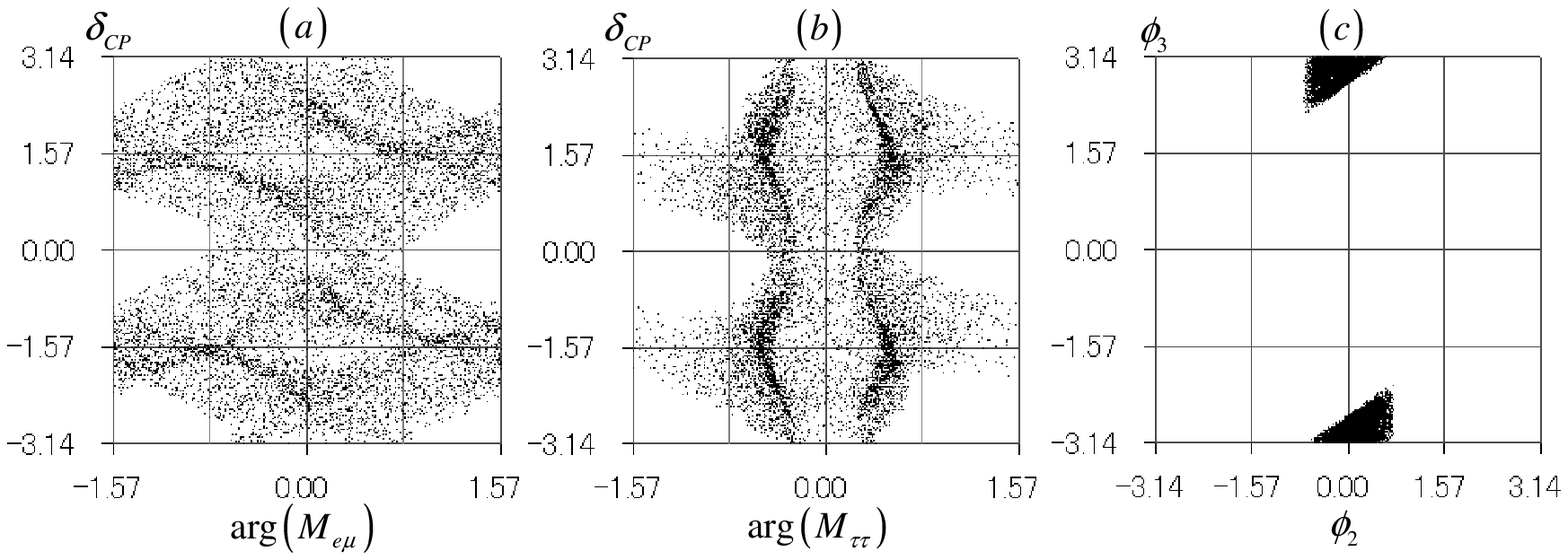}
\includegraphics*[10mm,226mm][300mm,290mm]{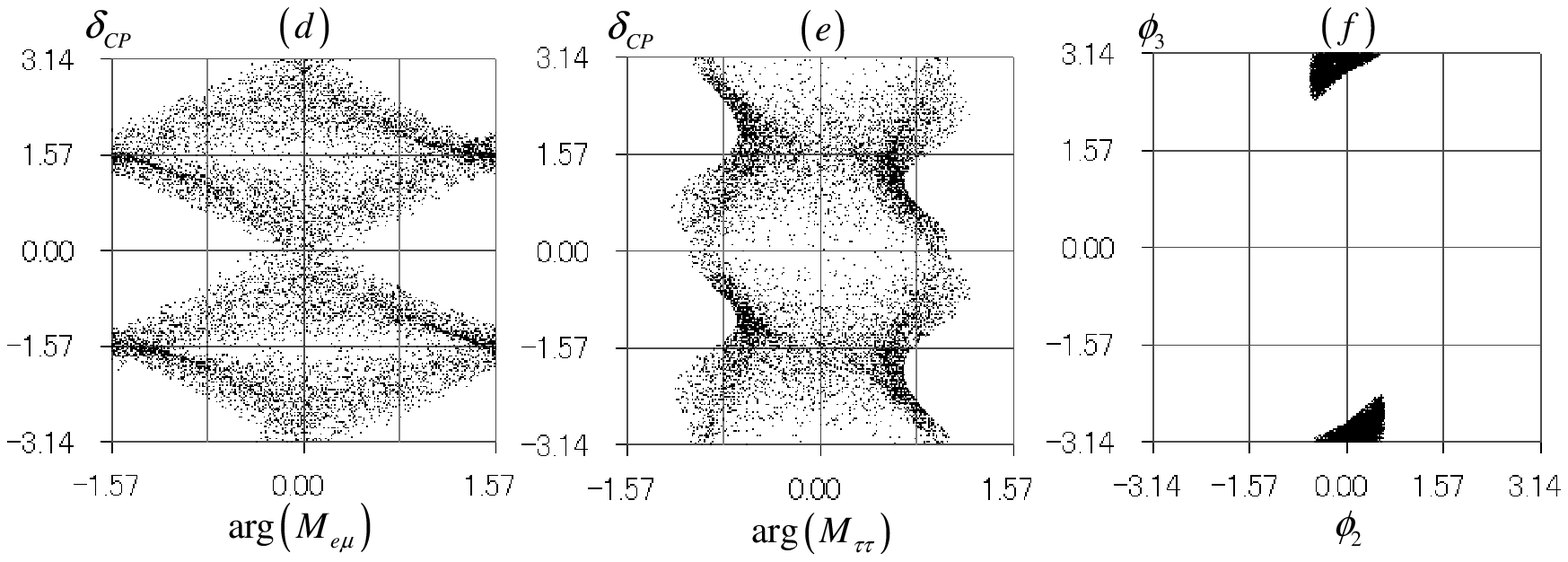}
\caption{The same as in FIG.\ref{Fig:Normal} for (a) and (b) but (c) $\phi_3$ as a function of $\phi_2$ for the degenerate mass pattern with ${\tilde m}_1\approx {\tilde m}_2\approx -{\tilde m}_3$ (DMP) satisfying the normal mass ordering.  Lower figures (d)-(f) show results for the inverted mass ordering.}
\label{Fig:Degenerate}
\end{center}
\end{figure}

We start with the following neutrino mass matrix $M_\nu$, which has three complex flavor neutrino masses
$M_{e\mu}$, $M_{e\tau}$ and $M_{\tau\tau}$.  This choice of phases is suggested by Eq.(\ref{Eq:Msin13zero}) and yields
\begin{eqnarray}
&&
M_\nu=
\left( {\begin{array}{*{20}{c}}
   \left| M_{ee}\right|  & M_{e\mu} & M_{e\tau} \\
   M_{e\mu} & \left| M_{\mu\mu}\right|  & \left| M_{\mu\tau} \right|  \\
   M_{e\tau} & \left| M_{\mu\tau}\right|  & M_{\tau\tau}  \\
\end{array}} \right).
\label{Eq:Mnu-start}
\end{eqnarray}
The mass matrix $M$ physically equivalent to $M_\nu$ can be obtained by including the freedom of three charged-lepton phases denoted by $\theta_{e,\mu,\tau}$ and is expressed to be:
\begin{eqnarray}
&&
M=
\left( {\begin{array}{*{20}{c}}
{{e^{ - 2i{\theta_e}}}\left| {{M_{ee}}} \right|}&{{e^{ - i\left( {{\theta_e} + {\theta_\mu }} \right)}}{M_{e\mu }}}&{ - {e^{ - i\left( {{\theta_e} + {\theta_\tau }} \right)}}{M_{e\tau }}}\\
{{e^{ - i\left( {{\theta_e} + {\theta_\mu }} \right)}}{M_{e\mu }}}&{{e^{ - 2i{\theta_\mu }}}\left| {{M_{\mu \mu }}} \right|}&{{e^{ - i\left( {{\theta_\mu } + {\theta_\tau }} \right)}}\left| {{M_{\mu \tau }}} \right|}\\
{ - {e^{ - i\left( {{\theta_e} + {\theta_\tau }} \right)}}{M_{e\tau }}}&{{e^{ - i\left( {{\theta_\mu } + {\theta_\tau }} \right)}}\left| {{M_{\mu \tau }}} \right|}&{{e^{ - 2i{\theta_\tau }}}{M_{\tau \tau }}}
\end{array}} \right).
\label{Eq:Mnu}
\end{eqnarray}
One has to diagonalize Eq.(\ref{Eq:Mnu}) to give $m_{1,2,3}$.  Since Eq.(\ref{Eq:Mnu}) contains six phases associated with six complex masses, the relevant $U_{PMNS}$, $U^\prime_{PMNS}$, should contain six phases, among which three phases are redundant \cite{generalU1,generalU1-2,generalU2}. We use three phases denoted by $\delta$ associated with the 1-3 mixing, $\gamma$ associated with the 2-3 mixing and $\rho$ associated with the 1-2 mixing and another three phases denoted by $\alpha_{1,2,3}$ as Majorana phases to define $U^\prime_{PMNS}$ \cite{generalU1}.  After three redundant phases $\rho$, $\gamma$ and $\varphi_1$ are removed from $U^\prime_{PMNS}$, Eq.(\ref{Eq:Mnu}) is modified into:
\begin{eqnarray}
&&
M^{PDG}=
\left( {\begin{array}{*{20}{c}}
{{e^{2i{\rho _e}}}\left| {{M_{ee}}} \right|}&{{e^{i\left( {{\rho _e} + {\gamma _\mu }} \right)}}{M_{e\mu }}}&{{e^{i\left( {{\rho _e} - {\gamma _\tau }} \right)}}{M_{e\tau }}}\\
{{e^{i\left( {{\rho _e} + {\gamma _\mu }} \right)}}{M_{e\mu }}}&{{e^{2i{\gamma _\mu }}}\left| {{M_{\mu \mu }}} \right|}&{{e^{ i\left( {{\gamma _\mu }-{\gamma _\tau }} \right)}}\left| {{M_{\mu \tau }}} \right|}\\
{{e^{i\left( {{\rho _e} - {\gamma _\tau }} \right)}}{M_{e\tau }}}&{{e^{ i\left( {{\gamma _\mu }-{\gamma _\tau }} \right)}}\left| {{M_{\mu \tau }}} \right|}&{{e^{ - 2i{\gamma _\tau }}}{M_{\tau \tau }}}
\end{array}} \right),
\label{Eq:MPDGnu}
\end{eqnarray}
where ${\rho _e} = \rho  - {\theta _e}$, ${\gamma _\mu } = \gamma  - {\theta _\mu }$ and ${\gamma _\tau } = \gamma  + {\theta _\tau }$, which can be diagonalized by $U_{PMNS}$ of Eq.(\ref{Eq:UuPDG}) with $\delta_{CP} = \delta+\rho$, $\phi_2=\varphi_2-\varphi_1$ and $\phi_3=\varphi_3-\varphi_1$ for  $\varphi_1 = \alpha_1-\rho$ and $\varphi_{2,3} = \alpha_{2,3}$.  As a result, phases of $M_{e\mu}$, $M_{e\tau}$ and $M_{\tau\tau}$ are expressed in terms of $\arg \left( M^{PDG}_{ij} \right)$ ($i,j$ =$e$, $\mu$, $\tau$) as follows: 
\begin{eqnarray}
\arg \left( {{M_{e\mu }}} \right) &=&\arg \left( {M_{e\mu }^{PDG}} \right) - \frac{{\arg \left( {M_{ee}^{PDG}} \right) + \arg \left( {M_{\mu \mu }^{PDG}} \right)}}{2},
\nonumber\\
\arg \left( {{M_{e\tau }}} \right) &=& \arg \left( {M_{e\tau }^{PDG}} \right) - \frac{{\arg \left( {M_{ee}^{PDG}} \right) - \arg \left( {M_{\mu \mu }^{PDG}} \right)}}{2},
\label{Eq:Marg}\\
\arg \left( {{M_{\tau \tau }}} \right) &=& \arg \left( {M_{\tau \tau }^{PDG}} \right) + \arg \left( {M_{\mu \mu }^{PDG}} \right) - 2\arg \left( {M_{\mu \tau }^{PDG}} \right).
\nonumber
\end{eqnarray}

There is an alternative CP violation \cite{ThreeDiracPhases} induced by three CP-violating Dirac phases but without explicitly referring to Majorana phases.  For the 2-3 mixing, it uses an analogous Dirac phase to $\delta$ instead of $\gamma$, which is denoted by $\tau$ \cite{generalU1-2}, and $\tau$ is introduced as the same way as $\rho$ is. This parameterization denoted by $U_{PMNS}^{RV}$ is known to have an advantage to discuss property of $M^{PDG}_{ee}$ to be measured in $(\beta\beta)_{0\nu}$-decay \cite{DoubleBeta}, which is given by
\begin{equation}
M_{ee}^{PDG} = {e^{ - i{\varphi _1}}}\left[ {c_{13}^2\left( {c_{12}^2{m_1} + s_{12}^2{m_2}{e^{ - 2i\rho }}} \right) + s_{13}^2m_3{e^{2i\delta }}} \right].
\label{Eq:MeePDG}
\end{equation}
All three CP-violating Dirac phases are physical and observable and are related to $\delta_{CP}$ and ${\phi _{2,3}}$ as $\delta_{CP} = \delta+\rho+\tau$, ${\phi _2} = 2\rho$ and ${\phi _3} = 2\left( {\rho  + \tau } \right)$ leading to
\begin{eqnarray}
\delta  = {\delta _{CP}} - \frac{{{\phi _3}}}{2},
\quad
\rho  = \frac{{{\phi _2}}}{2},
\quad
\tau  = \frac{{{\phi _3} - {\phi _2}}}{2}.
\label{Eq:AllPhysical}
\end{eqnarray}
If $m_1=0$, CP-violating Majorana phase is $\varphi _3 - \varphi _2(= \phi)$ and $\tau$ and $\delta  + \rho$ are determined to be 
\begin{eqnarray}
\delta  + \rho = \delta _{CP} - \frac{\phi}{2},
\quad
\tau  = \frac{\phi}{2},
\label{Eq:AllPhysical_m1=0}
\end{eqnarray}
where $M_{ee}^{PDG}$ only depends on $\delta  + \rho$, while if $m_3=0$, CP-violating Majorana phase is $\phi _2$ and  $\rho$ and $\delta  + \tau$ are determined to be 
\begin{eqnarray}
\delta  + \tau  = \delta _{CP} - \frac{\phi _2}{2}. 
\quad
\rho  = \frac{\phi _2}{2},
\label{Eq:AllPhysical_m3=0}
\end{eqnarray}
where $M_{ee}^{PDG}$ only depends on $\rho$.

Our results of numerical calculations are listed in FIG.\ref{Fig:Simplified_normal}-FIG.\ref{Fig:Degenerate}. Shown in FIG.\ref{Fig:Simplified_normal} and FIG.\ref{Fig:Simplified_inverted} are predictions on $\delta_{CP}$ from the simplified relations Eqs.(\ref{Eq:deltaCP_PDG_normal}) and (\ref{Eq:deltaCP_PDG_inverted}). In the remaining figures, FIG.\ref{Fig:Normal}-FIG.\ref{Fig:Degenerate}, each of which corresponds to each mass pattern, predictions on $\delta_{CP}$ are depicted as functions of either $\arg\left(M_{e\mu}\right)$ or $\arg\left(M_{\tau\tau}\right)$, which exhibit a certain correlation with $\delta_{CP}$. On the other hand, any predominant correlation between $\delta_{CP}$ and $\arg\left(M_{e\tau}\right)$ in each case cannot be found.  Other correlations with CP-violating Majorana phases are also shown in the figures.  For the sake of simplicity, calculations have been done for $m_1=0$ eV for NMH and $m_3=0$ eV for IMH-1 and IMH-2.  For DMP, $m_1$=0.1 eV ($m_3$=0.1 eV) for the normal (inverted) mass ordering is adopted.  The parameters used are
\begin{eqnarray}
\Delta m^2_{21} ~[10^{-5}~{\rm eV}^2] & = &7.62,
\qquad
\Delta m^2_{31} ~[10^{-3}~{\rm eV}^2] = 2.53,
\label{NumericalMassSquared}\\
\sin ^2 \theta _{12} & = &0.32,
\qquad
\sin ^2 \theta _{23} =0.45,
\qquad
\sin ^2 \theta _{13} =0.025.
\label{NumericalAngle}
\end{eqnarray}
In Ref.\cite{NuData}, the suggested best fit value of $\delta_{CP}/\pi$ is 0.80 (-0.03) for normal (inverted) mass ordering although all values are allowed while, in Ref.\cite{NuData2}, the allowed region at the 1$\sigma$ range is 0.77-1.36 (0.83-1.47) for normal (inverted) mass ordering.

As can be seen from FIG.\ref{Fig:Simplified_normal} and FIG.\ref{Fig:Simplified_inverted}, it is observed that the figures indicate the approximate proportionality of $\delta_{CP}$ to predicted values of Eq.(\ref{Eq:deltaCP_PDG_normal}) for $\delta_{CP}$ using both $\arg\left( M_{e\mu}^{PDG}\right)$ and $\arg\left( M_{e\tau}^{PDG}\right)$ and of Eq.(\ref{Eq:deltaCP_PDG_inverted}) using $\arg\left( M_{e\tau}^{PDG}\right)$, which supports the validity of our predictions.  However, FIG.\ref{Fig:Simplified_inverted}-(a) for IMH shows that $\delta_{CP}$ using $\arg\left( M_{e\mu}^{PDG}\right)$ is not a suitable approximation and implies that the assumption of $\arg\left(M_{e\mu}^{PDG}\right)\approx \arg\left(M_{e\tau}^{PDG}\right)$ is not numerically supported. From Eqs.(\ref{Eq:deltaCP_PDG_normal}) and (\ref{Eq:deltaCP_PDG_inverted}), we obtain $\arg \left( M_{e\mu, e\tau}\right)$ related to $\delta_{CP}$ as
\begin{equation}
\arg \left( {{M_{e\mu }}} \right) \approx {\delta _{CP}} - \left[ {\arg \left( {M_{ee}^{PDG}} \right) - \arg \left( {M_{\mu \mu }^{PDG}} \right)} \right]/2,
\label{Eq:deltaCP_approx_normal}
\end{equation}
for NMH, and
\begin{equation}
\arg \left( {{M_{e\tau }}} \right) \approx  - {\delta _{CP}} + \left[ {\arg \left( {M_{ee}^{PDG}} \right) + \arg \left( {M_{\mu \mu }^{PDG}} \right)} \right]/2 + \pi,
\label{Eq:deltaCP_approx_inverted}
\end{equation}
for IMH-1. We have also checked that the approximated expressions of $\delta_{CP}$, Eqs.(\ref{Eq:deltaCP_normal})-(\ref{Eq:deltaCP_degenerate}), numerically well reproduce actual values of $\delta_{CP}$. 

For the results of $\arg\left( M_{e\mu,\tau\tau}\right)$ with respect to $\delta_{CP}$ and CP-violating Majorana phases, their suggested features are summarized as follows:
\begin{itemize}
\item For NMH with $m_1=0$, where $\phi=\varphi_3-\varphi_2$, FIG.\ref{Fig:Normal} indicates that
\begin{itemize}
\item $\delta _{CP}$ following the thick line approximated to be $\delta _{CP} \approx 2 \arg\left( M_{e\mu}\right)$ is realized by requiring $\arg \left( M_{ee}^{PDG}\right) - \arg \left( M_{\mu \mu }^{PDG}\right)\approx \delta_{CP}$ because of Eq.(\ref{Eq:deltaCP_approx_normal}),
\item $\left|\arg\left( M_{\tau\tau}\right)\right|\lesssim 0.5$,
\item $\phi$ has a simple dependence on $\arg\left( M_{\tau\tau}\right)$: $\phi=0,\pi$ if $\arg\left( M_{\tau\tau}\right)=0$ and $\phi=\pm\pi/2$ if $\vert\arg\left( M_{\tau\tau}\right)\vert$ reaches its maximal value of around 0.5.
\end{itemize}
\item For IMH-1, where $\phi=\varphi_2-\varphi_1$, FIG.\ref{Fig:Inverted1} indicates that
\begin{itemize}
\item $\delta_{CP}\rightarrow\pm\pi/2$ as $\arg\left( M_{e\mu}\right)\rightarrow\pm\pi/2$,
\item $\left|\arg\left( M_{\tau\tau}\right)\right|\lesssim 0.1$,
\item $\phi\approx 0$ is set by the condition of ${\tilde m}_1\approx {\tilde m}_2$.
\end{itemize}
\item For IMH-2, where $\phi=\varphi_2-\varphi_1$, FIG.\ref{Fig:Inverted2} indicates that
\begin{itemize}
\item $\pi/3\lesssim\left|\arg\left( M_{e\mu}\right)\right|\lesssim\pi/2$,
\item $\left|\arg\left( M_{\tau\tau}\right)\right|\lesssim 0.6$ for $\left| \delta_{CP}\right| \lesssim \pi/2$,
\item $\left|\arg\left( M_{\tau\tau}\right)\right|\lesssim 0.2$ if $\phi$ approaches toward $\pm\pi/2$.
\item $\phi\approx\pm\pi$ is set by the condition of ${\tilde m}_1 \approx -{\tilde m}_2$.
\end{itemize}
\item For DMP, FIG.\ref{Fig:Degenerate} indicates that
\begin{itemize}
\item $\pi/4\lesssim\left|\delta_{CP}\right|\lesssim 3\pi/4$ as $\arg\left( M_{e\mu}\right)\rightarrow\pm\pi/2$ for the normal mass ordering,
\item $\delta_{CP}\rightarrow\pm\pi/2$ as $\arg\left( M_{e\mu}\right)\rightarrow\pm\pi/2$ for the inverted mass ordering,
\item $\phi_2\approx 0$ and $\phi_3\approx \pm\pi$ for both mass orderings, which are linked to the fact that the sign of ${\tilde m}_3$ is different from that of ${\tilde m}_{1,2}$.
\end{itemize}
\end{itemize}
The phases of $M_{e\mu,\tau\tau}$ are taken to run from $-\pi/2$ to $\pi/2$.  It should be noted that $\arg\left(M_{e\mu}\right)$-$\delta_{CP}$ for IMH-1 (FIG.\ref{Fig:Inverted1}-(a)) and for DMP with the inverted mass ordering (FIG.\ref{Fig:Degenerate}-(a)) have the quite similar shape to each other, showing that the maximal Dirac CP-violation signalled by $\delta_{CP}= \pm \pi/2$ is realized by $\arg\left( M_{e\mu}\right)\approx \pm \pi/2$ and, at the same time, $\arg\left(M_{\tau\tau}\right)\approx 0$ is necessary for IMH-1.

\begin{figure}[t]
\begin{center}
\includegraphics*[10mm,226mm][300mm,290mm]{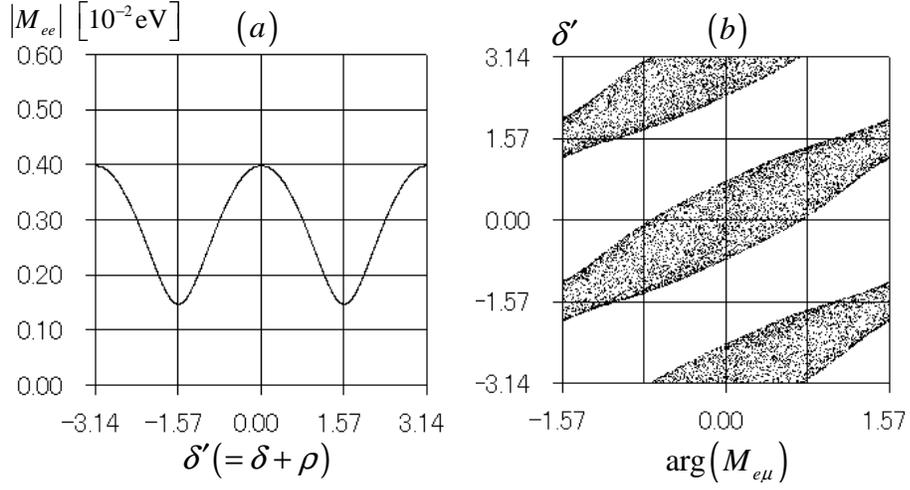}
\caption{The predictions of (a) $\left|M_{ee}\right|$ as a function of $\delta^\prime(=\delta+\rho)$ and (b) $\delta^\prime$ as a function of $\arg\left( M_{e\mu}\right)$ for NMH.}
\label{Fig:Normal-dirac}
\end{center}
\end{figure}
Also estimated is $\left| M_{ee}\right|(=\left| M^{PDG}_{ee}\right|)$ as the effective neutrino mass $m_{\beta\beta}$ in $(\beta\beta)_{0\nu}$-decay:
\begin{itemize}
\item $0.002 \lesssim\left| M_{ee}\right|~[{\rm eV}]\lesssim 0.004$ for NMH,
\item $\left| M_{ee}\right|~[{\rm eV}]\approx 0.05$ for IMH-1,
\item $0.02 \lesssim\left| M_{ee}\right|~[{\rm eV}]\lesssim 0.04$ for IMH-2,
\item $0.095 \lesssim\left| M_{ee}\right|~[{\rm eV}]\lesssim 0.1$ for DMP.
\end{itemize}
The results are consistent with naive estimation from Eqs.(\ref{M_angle_13=0_normal})-(\ref{M_angle_13=0_degenerate}).  Namely, the magnitude of $m_{\beta\beta}$ is suppressed for NMH.  To analyze $M^{PDG}_{ee}$ itself, it is useful to adopt $U_{PMNS}^{RV}$ parameterized by three Dirac phases, $\delta$ for the 1-3 mixing, $\rho$ for the 1-2 mixing and $\tau$ for the 2-3 mixing as have been already noted.  Differences between predictions by $U_{PMNS}^{RV}$ and those by $U_{PMNS}^{PDG}$ lie in the behavior of the CP-violating Majorana phases. Since these Majorana phases are constrained to be around 0 or $\pm\pi$ for IMH-1, IMH-2 and DMP, distinct differences cannot be expected.  Notable features in predictions by $U_{PMNS}^{RV}$ that we can observe are expected to arise for NMH.  Obvious one as shown in FIG.\ref{Fig:Normal-dirac} (a) is that $\left|M_{ee}\right|$ exhibits a clear correlation with $\delta^\prime (=\delta+\rho)$ for NMH as in Eq.(\ref{Eq:AllPhysical_m1=0}). Another one is shown in FIG.\ref{Fig:Normal-dirac} (b), where $M_{e\mu}$ and $\delta^\prime$ show a clear correlation that $\delta^\prime$ is scattered around the line $\delta^\prime=\arg\left(M_{e\mu}\right)$ (mod $\pi$). The corresponding prediction by $U_{PMNS}^{PDG}$ includes $\delta_{CP}$ as in FIG.\ref{Fig:Normal} (a) and shows that $\delta_{CP}$ is scattered in the entire region, which indicates no correlation with $\arg\left(M_{e\mu}\right)$ although the scattered points tend to form a straight line.

To summarize, we have derived a general formula to calculate $\delta_{CP}$ expressed in terms of the corrections $\delta M_{e\tau }^{PDG}$ and $\delta M_{\tau\tau}^{PDG}$ to neutrino mixings with $\theta_{13}=0$. The formula is given by Eq.(\ref{Eq:deltaCP_masssquared}):
\begin{equation}
{\delta _{CP}} \approx \arg \left[ {\left( {\frac{1}{{{t_{23}}}}M_{\mu \tau }^{PDG\ast } + M_{\mu \mu }^{PDG\ast }} \right)\delta M_{e\tau }^{PDG} + M_{ee}^{PDG}\delta M_{e\tau }^{PDG\ast } - {t_{23}}M_{e\mu }^{PDG}\delta M_{\tau \tau }^{PDG\ast }} \right],
\label{Eq:deltaCP_formula}
\end{equation}
where an extra $\pi$ should be added if $m_3^2 - \left(c_{12}^2m_1^2 + s_{12}^2m_2^2\right) < 0$.  These $\delta M_{e\tau }^{PDG}$ and $\delta M_{\tau\tau}^{PDG}$ are described in terms of $M^{PDG}$ as $\sin\theta_{13}\delta M^{PDG}_{e\tau} =M_{e\tau }^{PDG} + t_{23}M_{e\mu }^{PDG}$ and $\sin\theta_{13}\delta M^{PDG}_{\tau\tau} =M_{\tau \tau }^{PDG} - \left[ M_{\mu \mu }^{PDG} + \left(1 - t_{23}^2\right)M_{\mu \tau }^{PDG}/t_{23}\right]$. Their mass dependence is then determined to be:
\begin{eqnarray}
&&
\delta M^{PDG}_{e\tau} = \frac{{c_{13}}}{{c_{23}}}\left[ {{e^{ i{\delta _{CP}}}}{{\tilde m}_3} - {e^{-i{\delta _{CP}}}}\left( {c_{12}^2{{\tilde m}_1} + s_{12}^2{{\tilde m}_2}} \right)} \right],
\quad
\delta M^{PDG}_{\tau\tau} = \frac{{c_{12}}{s_{12}}}{{s_{23}}{c_{23}}}{e^{-i{\delta _{CP}}}}\left( {{{\tilde m}_2} - {{\tilde m}_1}} \right).
\label{Eq:deltaCPMajoranaPhase}
\end{eqnarray}
Other useful findings are
\begin{enumerate}
\item  The main source of $\delta_{CP}$ is $\delta M_{e\tau }^{PDG}$ except for IMH-2 because $\delta M_{\tau\tau }^{PDG}$ is not suppressed if ${\tilde m}_1\approx -{\tilde m}_2$, and
\item $\delta_{CP}$ is well predicted to be $\arg \left( {M_{e\mu }^{PDG}} \right) - \arg \left( {M_{\mu \mu }^{PDG}} \right)$ with $\arg \left( M_{e\tau}^{PDG} \right)\approx\arg \left(M_{e\mu }^{PDG} \right)$ for NMH and $\arg \left( {M_{ee}^{PDG}} \right) - \arg \left( {M_{e\tau }^{PDG}} \right) + \pi$ for IMH-1.
\end{enumerate}

For the specific neutrino masses, whose phases are adjusted to arise from $M_{e\mu,e\tau,\tau\tau}$, the effects of CP-violation caused by each flavor neutrino mass are expressed in terms of $M^{PDG}$ according to Eq.(\ref{Eq:Marg}).  For the numerical calculations, we adopted $m_1=0$ eV ($m_3=0$ eV) for NMH (IMH) and $m_1=0.1$ eV ($m_3=0.1$ eV) for DMP with the normal (inverted) mass ordering.  It is, then, numerically indicated that $\delta_{CP}$ tends to satisfy $\delta _{CP} \approx 2 \arg\left( M_{e\mu}\right)$ requiring the relation of $\arg \left( M_{ee}^{PDG}\right) - \arg \left( M_{\mu \mu }^{PDG}\right)\approx \delta_{CP}$ in NMH. In the inverted mass hierarchies, we have observed that $\left|\arg\left( M_{\tau\tau}\right)\right|\lesssim 0.1$ for IMH-1 and $\pi/3\lesssim \vert\arg\left( M_{e\mu}\right)\vert\lesssim \pi/2$ for IMH-2. CP-violating Majorana phase $\phi_2$ ($\phi_3$) for DMP is limited to locate around 0 ($\pm\pi$) owing the mass relation of ${\tilde m}_1 \approx {\tilde m}_2 \approx -{\tilde m}_3$. Effects of Majorana CP-violation are expected to be suppressed for DMP.  On the other hand, for NMH, Majorana CP-violation tends to be maximal as $\left|\arg\left( M_{\tau\tau}\right)\right|$ reaches its maximal value of $\approx 0.5$. If Majorana CP-violation tends to be maximal, we have also found that $\left|\arg\left( M_{\tau\tau}\right)\right|\lesssim 0.2$ for IMH-2.  Dirac CP-violation gets maximal as $\arg\left( M_{e\mu}\right)\rightarrow \pm \pi/2$ for IMH-1 and DMP with the inverted mass ordering and $\arg\left( M_{\tau\tau}\right)\approx0$ is also satisfied for IMH-1. 

Another parameterization of $U_{PMNS}$ utilizes three CP-violating Dirac phases $\delta$, $\rho$, and $\tau$, where the CP-violating phases in the PDG version are determined to be $\delta_{CP} = \delta+\rho+\tau$, ${\phi _2} = 2\rho$ and ${\phi _3} = 2\left( {\rho  + \tau } \right)$.  There are some advantages of choosing $U^{RV}_{PMNS}$ over $U^{PDG}_{PMNS}$ found in the present analysis: 
\begin{enumerate}
\item The oscillation behavior of $\left|M_{ee}\right|$ is well traced for NMH as already pointed out \cite{ThreeDiracPhases} and is useful to determine $\delta^\prime(=\delta+\rho)$ from $\left|M_{ee}\right|$;
\item In NMH, $\delta^\prime$ is scattered around the line of $\delta^\prime=\arg(M_{e\mu})$ (mod $\pi/2$)  while $\delta_{CP}$ is scattered in the entire region.
\end{enumerate}
It is in principle possible to know an allowed range of $\arg(M_{e\mu})$ from $\delta^\prime$ to be extracted from  $\left|M_{ee}\right|$ if it is measured.  To say something more about the alternative CP-violation for NMH as well as IMH-1 and IMH-2, we have to include effects of two active CP-violating Majorana phases associated with three nonvanishing neutrino masses and results of CP-violation will be discussed elsewhere. 

\vspace{3mm}
\noindent

\centerline{\small \bf ACKNOWLEGMENTS}

The author is grateful to T. Kitabayashi for reading manuscript and useful comments.


\end{document}